\documentstyle[12pt,fleqn]{article}
\def\fnote#1#2{\begingroup\def\thefootnote{#1}\footnote{#2}\endgroup}
\makeatletter
\def\section{\@startsection {section}{1}{\z@}{3.5ex plus 1ex minus
    .2ex}{2.3ex plus .2ex}{\sc }}
\def\subsection{\@startsection{subsection}{2}{\z@}{3.25ex plus 1ex
minus
   .2ex}{1.5ex plus .2ex}{\small \sc }}
\def\appendix{\par\clearpage
  \setcounter{section}{0}
  \setcounter{subsection}{0}
  \@addtoreset{equation}{section}
  \def\@sectname{Appendix~}
  \def\theequation{\thesection.\arabic{equation}}
  \def\thesection{\Alph{section}}}
\makeatother
\makeatletter \@addtoreset{equation}{section} \makeatother
\renewcommand{\theequation}{\thesection.\arabic{equation}}

\def\ap#1#2#3{     {\it Ann. Phys. (NY) }{\bf #1} (19#2) #3}

\def\npb#1#2#3{    {\it Nucl. Phys. }{\bf B #1} (19#2) #3}
\def\plb#1#2#3{    {\it Phys. Lett. }{\bf B #1} (19#2) #3}
\def\prd#1#2#3{    {\it Phys. Rev. }{\bf D #1} (19#2) #3}

\def\prl#1#2#3{    {\it Phys. Rev. Lett. }{\bf #1} (19#2) #3}

\def\zpc#1#2#3{    {\it Z. Physik }{\bf C #1} (19#2) #3}
\def\mpla#1#2#3{   {\it Mod. Phys. Lett. }{\bf A #1} (19#2) #3}
\def\sjnp#1#2#3{{\it Sov. J. Nucl. Phys. }{\bf #1} (19#2) #3}

\def\nc#1#2#3{     {\it Nuovo Cim. }{\bf #1} (19#2) #3}

\def\ijmpa#1#2#3{  {\it Int. J. Mod. Phys. }{\bf A #1} (19#2) #3}
\def\ltap{\ \raisebox{-.4ex}{\rlap{$\sim$}} \raisebox{.4ex}{$<$}\ }
\def\gtap{\ \raisebox{-.4ex}{\rlap{$\sim$}} \raisebox{.4ex}{$>$}\ }
\def\eq#1{{eq.~(\ref{#1})}}

\def\Im{\mathop{\mbox{Im}}}

\def\etal{{\it et al.}}

\newcommand{\bea}{\begin{eqnarray}}
\newcommand{\beq}{\begin{equation}}
\newcommand{\eea}{\end{eqnarray}}
\newcommand{\eeq}{\end{equation}}
\newcommand{\nnu}{\nonumber}
\newcommand{\spav}[1]{\parbox{1mm}{\vspace*{#1}}}

\def\be{\begin{equation}}
\def\ee{\end{equation}}
\def\ba{\begin{array}}
\def\ea{\end{array}}
\def\l{\label}

\def\LR{{\rm \scriptscriptstyle LR}}

\def\refe#1{(\ref{#1})}

\def\ltap{\ \raisebox{-.4ex}{\rlap{$\sim$}} \raisebox{.4ex}{$<$}\ }
\def\gtap{\ \raisebox{-.4ex}{\rlap{$\sim$}} \raisebox{.4ex}{$>$}\ }
\def\etal{{\it et al.}}

\def\pt{{\cal P}_\perp}
\def\kmu3{{$K^+\to \pi^0 \mu^+ \nu_\mu$}}
\def\k0mu3{{$K^0_{\mu3}$}}
\def\bv{{\vec{\beta}}}

\begin{document}

\begin{titlepage}
\spav{3cm}

\begin{center}

{\Large\bf The supersymmetric prediction}\\ 
{\Large\bf for the muon transverse polarization}\\
{\Large\bf in the \kmu3\ decay}\\
\spav{1.8cm}\\
 {\large Marco Fabbrichesi$^{a,b}$ and Francesco Vissani$^{a,c}$ }
\spav{1cm}\\
{\em ${}^a$ Istituto Nazionale di Fisica Nucleare, 
\\[-1ex] Sezione di Trieste}\\
{\em ${}^b$ Scuola Internazionale Superiore di Studi Avanzati,
\\[-1ex] Via Beirut 2-4, I-34013 Trieste, Italy} \\
{\em ${}^c$ International Centre for Theoretical Physics, 
\\[-1ex] Strada Costiera 11, I-34013 Trieste, Italy}
\end{center}

\noindent
\hrulefill

{\small 
\noindent 
{\sc \bf Abstract}\\
\noindent
The muon transverse polarization in the  \kmu3\ decay 
will be measured  at the $10^{-4}$ level in forthcoming 
experiments.
We compare the phenomenological perspectives 
with the theoretical predictions in
supersymmetric extensions
of the standard model.
In the minimal extension, 
CP-violating phases lead to a non-zero transverse polarization,
that however is too small to account for a 
positive experimental signal.
The problems that one encounters when departing 
from minimal assumptions are discussed. 
An observable effect is possible if the hypothesis 
of R-parity conservation is relaxed, 
but only at the price of assuming
a very special pattern for the R-parity breaking couplings.} 

\noindent
\hrulefill

\vfill

\flushleft{\tt IC/96/220\\[-0.28ex]
SISSA 140/96/EP\\[-0.28ex]
November 1996}

\end{titlepage}

\newpage 
\setcounter{footnote}{0}
\setcounter{page}{1}

A non-vanishing component of the muon polarization, 
transverse to the decay plane of the \kmu3\ (\k0mu3) 
process, would signal CP-violating effects~\cite{Sakurai}, if 
bigger than  $\pt \sim 10^{-6}$ (the  contribution
due to final state interactions~\cite{Zhitniskii}).
Such a signal would be particularly interesting, 
since it would imply physics beyond the 
standard model~\cite{Golowich}. 

Forthcoming experiments at DA$\Phi$NE~\cite{DAFNE}, 
KEK~\cite{Kuno} 
and BNL \cite{BNL} may push the present limit 
$\pt<5\times 10^{-3}$~\cite{old} by 
more than one order of magnitude: 
$\pt<2.8\times 10^{-4}$ 
(from \cite{Privitera}), 
or perhaps obtain a positive result.
For this reason, it is important to state the prediction for this 
observable in all those  models that are potential
candidates to describe physics
beyond the standard model 
(see \cite{Zhitniskii,Cheng,Leurer,CFK,gk,bg} 
for earlier studies).

The outline of the present work is the following. 
Sect.\ 1 is devoted to a phenomenological discussion, including
the relationship between the polarization 
and the invariant form factors,
the discussion
of the experimental perspectives and the specification of 
the effect of strong interactions.
We analyze in Sect.\ 2 the size of the transverse
polarization in possible extensions of the standard model. 
We focus in particular on supersymmetric models 
(see \cite{CF} and \cite{WuNg} for previous analyses):
the minimal extension of the standard model, 
and models with explicit R-parity breaking.
The last Section is devoted to the conclusions.

\section{Decay Form Factors and Muon Polarization}

The form of the invariant amplitude 
suggested by the Particle Data Group  
is~\cite{PDG}:
\be
\ba{rl}
{\cal M}=\displaystyle {G_F\sin\theta_C}\ 
\overline{u(p_\nu)_L} \
[& \!\left( F_+ \ (p_K +p_\pi)^a - 
F_-  \ (p_K -p_\pi)^a \right)\ \gamma_a 
\\[1ex]
& \!\displaystyle + 2 F_S \ m_K 
+ 2i \frac{F_T}{m_K} \ \sigma_{ab}\ p_K^a p_\pi^b \ ] \
v(p_\mu,\vec{s}_\mu) .
\ea
\l{trad}
\ee
The leptons in the final state are a
left-handed neutrino with four-momentum $p_\nu,$ and
an antimuon with four-momentum $p_\mu$ and spin $\vec{s}_\mu$ 
(Notice that, due to the chirality projector, we can account for
an additional contribution 
$\delta F_T/m_K\  p_K^a p_\pi^b \sigma^{cd}\epsilon_{abcd} $ 
to the bracketed term simply by redefining $F_T\to F_T-\delta F_T$).

In \eq{trad} we  distinguish between the contribution 
of the usual $V-A$  interactions and 
that of other possible interactions because
 final state interactions \cite{Zhitniskii}
and the standard model sources of CP violation 
\cite{Golowich} give a negligible
contribution to $F_S$ and $F_T$. 

The form factors $F_+$, $F_-$, $F_S$, $F_T$ in \eq{trad}
depend in general on the hadronic momentum 
transferred, $q^2=(p_K-p_\pi)^2$;
CP invariance implies that they are relatively real.
They can be calculated once the model
and the hadronic matrix elements are specified. 

\subsection{The Transverse Polarization}

By the use of the equations of motions we can recast  \refe{trad}
in the form:
\be
{\cal M}=2 G_F  \sin\theta_C \ m_K\ 
F_+\ (1 - \xi_T\ \frac{m_\mu}{m_K})\ 
\overline{u(p_\nu)_L} \
\left[\ \frac{{p}_K^a}{m_K}\ \gamma_a + \zeta\ \right]
v(p_\mu,\vec{s}_\mu) .
\l{simple}
\ee 
The amplitude for scalar interactions is proportional to
\beq
\!\!\!\! \zeta  =  \frac{1}{2}\left[ \frac{m_\mu}{m_K} 
 + \left(
\frac{m_\mu}{m_K}\ \xi 
+ 2 \, \xi_S  
+  \frac{2\, p_K \cdot (p_\nu-p_\mu) + m_\mu^2}{m_K^2}\ \xi_T \right)
 \left( 1 -  \frac{m_\mu}{m_K}\ \xi_T \right)^{-1}
\right] \l{zeta}
\eeq
where we defined 
\begin{equation}
\xi=\frac{F_-}{F_+},\ \ \ \ \ \xi_S=\frac{F_S}{F_+}\ \ \ \ \ 
{\rm and}\ \ \ \ \ \xi_T=\frac{F_T}{F_+}\, .\label{xi}
\end{equation} 
We assume that the parameters $\xi,\xi_S,\xi_T$ 
are constants. Actually, writing
eq.~\refe{trad} with four form factors makes sense {\em only 
if} their dependence on the invariants is to a certain 
extent specified: 
Lorentz invariance alone would require just two form factors.

The form \refe{simple} 
is very convenient in evaluating the 
decay rate (this was originally pointed out in \cite{OK}). 
In bispinorial notations we can write:
\be
{\cal M}\propto \sqrt{2 E_\nu\ 2 E_\mu}\  
\phi(-\bv_\nu)^\dagger \ [\ b_+ + b_-\ \bv_\mu\cdot \vec{\sigma}\ ]\
\phi(-\vec{s}_\mu),
\l{ampbisp}
\ee
where $\bv$ are the velocity vectors for 
the neutrino and the antimuon,
and $\gamma_\mu=1/\sqrt{1-\beta^2_\mu}$ 
the antimuon Lorentz boost factor
(not a gamma matrix!). The spinors $\phi(\vec{n})$ obey 
$\phi^\dagger \phi=1$ and 
$(\vec{n}\vec{\sigma})\phi(\vec{n})=\phi(\vec{n}).$ 
The two coefficients 
of the amplitude can be expressed 
as $b_\pm=(\zeta\mp 1) (1+\gamma_\mu^{-1})^{\pm 1/2}.$  
All quantities are evaluated in the kaon rest frame.
Squaring \refe{ampbisp} we obtain:
\be
|{\cal M}|^2 \propto 8\ E_\nu E_\mu\ {\cal P}\ 
\cos^2[\Theta/2],
\l{square}
\ee 
where $\Theta$ is the angle between the 
observed polarization $\vec{s}_\mu$
and the vector $\vec{\cal P},$
\be
\ba{rcl}
\vec{\cal P} &=&
[\ |1-\zeta|^2\  \gamma_\mu^{-1}-
2{\rm Re}\,\zeta\ (1-\gamma_\mu^{-1})\ ]
\ \bv_\nu\\
&+&[\ 1-|\zeta|^2 + |1+\zeta|^2\  \bv_\nu \cdot \bv_\mu\
(1+\gamma_\mu^{-1})\ ]\ \bv_\mu\\ 
&+& {\rm Im}\,\zeta \left[\ \bv_\nu \times \bv_\mu\ \right] \, ,
\ea
\l{p}
\ee
whose modulus, which enters formula \refe{square} is:
\be
{\cal P}=1+\bv_\nu \cdot \bv_\mu
-2 {\rm Re}\,\zeta\ \gamma_\mu^{-1} 
+ |\zeta|^2\ (1-\bv_\nu \cdot \bv_\mu) .
\l{pmod}
\ee
It is worthwile emphasize some aspects of this result,
which was originally derived in \cite{pt1,pt2}:
\begin{description}
\item (a) The probability of transition, 
proportional to $|{\cal M}|^2$,
is small for small lepton energies.
\item (b) The three terms in equation \refe{pmod}
have a clear interpretation:
the first (last) describes 
left-(right-)handed particle, produced by vector
(scalar) interactions, and the second is a typical 
interference term 
($\zeta$ parameterizes the relative amount of 
scalar-like interactions).
\item (c) The squared cosine  factor in \refe{square} indicates
that the antimuon produced in the decay is completely polarized 
along $\vec{\cal P}.$ 
The non-polarized case is recovered upon averaging (according to
$\cos^2[\Theta/2]\to 1/2)$.
\item (d) For slow antimuons, formula \refe{p} shows
that the polarization is parallel to the neutrino velocity
vector.
This is clearly due to the fact that the neutrino is
in a negative helicity state. 
\item (e) The transverse polarization, the last term in 
\refe{p}, is maximum when the antimuon and the neutrino velocities 
are orthogonal in the laboratory frame.
\end{description}

\subsection{Phenomenological Remarks and Experimental Perspectives}

In the Dalitz plot 
distributions, the imaginary part of the parameter
$\zeta$ in eq.\ \refe{zeta}  (or equivalently $\xi$ and/or 
$\xi_S$ and/or $\xi_T$ in eq.\ \refe{xi})
enters only quadratically (compare with eq.\ \refe{pmod}).
On the other hand, the dependence of \refe{p} on 
$\Im\,\zeta$ is linear and 
therefore the transverse
polarization would give to a large 
extent an independent information on
the form factors.

Recently 
a simple experimental method has been proposed \cite{Privitera}
to detect the effects 
of the polarizations at DA$\Phi$NE without the 
need of a polarimeter. 
It is based on the fact that the direction of the positron emitted  
in the antimuon decay  is correlated to $\vec{\cal P},$
and therefore to the kinematical variables 
of the decay that produced the antimuon.
There is a different probability
of emission above and below the
decay plane if $\zeta$ 
has an imaginary part
(the direction ``above 
the decay plane'' is specified by the 
vector $\beta_\nu\times\beta_\mu;$
no definition is needed when the particles are collinear). 
Recalling that the differential probability 
of emission of a positron
with angle $\theta,$ measured from 
the antimuon polarization direction, is
$(1+1/3\cos\theta)\ d\Omega/4\pi,$ 
where $\Omega$ is the solid angle, we find 
$P({\mathrm above})-P({\mathrm below})={\cal P}_\perp/(6\ {\cal P}),$ 
where ${\cal P}_\perp=\Im\zeta\ |\bv_\nu\times\bv_\mu|.$ Therefore
the asymmetry rate is:
\be
\Im\zeta\ \frac{G_F^2 \sin^2\theta_C}{48\pi^3} \
m_K \left| F_+\ (1 - \xi_T\frac{m_\mu}{m_K}) \right| ^2 \
\left| \vec{p}_\nu \times \vec{p}_\mu \right| \ dE_\nu dE_\mu.
\ee

Notice that, according to eqs.\ \refe{zeta} and \refe{p},
if the leading contribution to the transverse polarization 
comes from ${\rm Im}\,\xi$ or ${\rm Im}\,\xi_S,$ 
then ${\cal P}_\perp$ 
points in a given half-space 
with respect to the decay plane; 
on the contrary, 
if  ${\cal P}_\perp$ is related to ${\rm Im}\,\xi_T$ 
it can point in both directions because 
the factor $\left( E_\nu-E_\mu + m_\mu^2/m_K \right)$
changes sign (see eq.\ \refe{zeta}).

In this way, DA$\Phi$NE can provide 
a factor ten improvement of
the current limits, and therefore
may reveal transverse polarization effects if~\fnote{\dag}{See 
ref.~\cite{Privitera}.
Note however that the different dependence on the
energy in eq.\ \refe{zeta} implied by the tensor form factor 
requires a different experimental analysis from 
the case in which one assumes a pure scalar contribution
(the one considered in \cite{Privitera}).}:
\be
\Im \xi_S\ \ \ {\rm or}\ \ \ \Im \xi_T \ \gtap 2\times 10^{-3} . 
\label{obs}
\ee
It is suggestive to
compare this figure with the most 
recent experimental analysis
of $K^0_{e3}$ decays \cite{Akimenko}, for which
the best fit of the Dalitz-plot distributions requires
$F_S=(7\pm 1.6\pm 1.6)\times 10^{-2}$,
$F_T=(5.3\pm 1.0\pm 1.0)\times 10^{-1}.$
If similar values for $K^0_{\mu3}$ decay parameters
are assumed, an imaginary part as small as 0.5 \% 
would lead to an observable signal.

\subsection{Current Form Factors}

The Lorentz invariant decomposition of the hadronic 
matrix elements introduces five form factors:
\bea
\langle \pi^0 | \bar{s}(0) \gamma^a u(0) | K^+\rangle & = &
\frac{1}{\sqrt{2}}\ [\ f_+(q^2)\ (p_K+p_\pi)^a  - f_-(q^2)\ 
(p_K-p_\pi)^a \ ] \\
\langle \pi^0 | \bar{s}(0) u(0) | K^+\rangle & = &
{\sqrt{2}\ m_K}\ f_S(q^2)  \\
\langle \pi^0 | \bar{s}(0) \sigma^{ab} u(0) | K^+\rangle &  =&
\frac{i}{\sqrt{2}\ m_K}\ 
f_T(q^2)\ (p_K^a p_\pi^b - p_K^b p_\pi^a) \nnu \\   
    & &                + f_T'(q^2)\ \epsilon^{abcd} (p_K)_a(p_\pi)_b
\eea
The specification of these form factors amounts to the
description of the nonperturbative effects of strong interactions.
Renormalization group factors induced by QCD 
are of order unity for such
a semi-leptonic decay and we shall ignore them.
At tree level, $V-A$ structure entails 
only the first operator.
The other two are in general present 
when new interactions are introduced.

The flavor $SU(3)$ symmetry imposes relations 
on the vector form factors so that
$f_+(0)=-1$ and $f_-(q^2)=0$
(see for instance the discussion in \cite{OkunBook}). 
For the purpose of describing the transverse polarization,
it is adequate to consider such a  $SU(3)$ symmetric limit.\newline
It is important to point out that the precise 
knowledge of the form factors, and in particular of their 
momentum dependence is important in {\em testing} the hypothesis 
of pure standard model 
interactions when studying the $K_{\mu3}$ Dalitz plot 
distributions.

The scalar form factor can be computed 
considering the matrix element of the divergence of
the vector current, and then using the free equations for
the quark fields (as described in \cite{Cheng}):
\be
f_S= \frac{(m_{K}^2-m_{\pi}^2) f_+ + q^2 f_-}{2\ m_K\ (m_s-m_u)}
\simeq \frac{m_K}{2\ m_s} f_+
\l{scalar}
\ee
This estimation
is subject to considerable uncertainty due to the use 
of the free equations of motion and the value of $m_s$
(we use the central value of the recent determination
$m_s\equiv \bar m_s (1 \: \mbox{GeV}) = 175 \pm 25$ MeV~\cite{BP}). 
Nevertheless, it is sufficiently accurate for the purpose 
of estimating the muon transverse polarization. 

For completeness, we also present the estimation 
of the tensor form factors, that has been first 
evaluated in~\cite{Chizhov}.
In the context of the 
chiral quark model~\cite{QM}, at the leading order, we find: 
\be
f_T\simeq \frac{6\ m_K\ M}{\Lambda_\chi^2}\approx 1,
\ \ \ \ f_T'\approx 0 .
\ee
where $M$ is the constituent quark mass,
a model dependent parameter ($M=220$ MeV in the above), and
$\Lambda_\chi\approx 1$ GeV is the chiral symmetry breaking scale.

\section{The Theoretical Prediction}

Which kind of models can lead to a non-zero 
transverse polarization, and
what are their phenomenological implications?
This question will be discussed 
within the supersymmetric extensions of the standard model. 
We first examine the minimal extension, 
where the effect arises at the one-loop level, 
and then consider R-parity 
non-conserving interactions, that 
can give an effect already at the tree level.

\subsection{Minimal Supersymmetric Standard Model}
\vskip.3truecm 
{\it Introduction}
\vskip.3truecm 
The supersymmetrization of the standard model requires
enlarging the spectrum of the theory. 
The squarks and the sleptons are the scalar partner of the
quarks and leptons; 
two Higgs
doublets are present, each one paired with a fermionic doublet
(higgsino);
similarly, for each gauge boson one
fermionic degree of freedom appears (gaugino).
According to the usual convention we denote
the supersymmetric particles by a tilde: for instance $\tilde g$
is the gluino, $\tilde q$ a generic squark.
The values of the supersymmetric partners masses
depend on the mechanism of supersymmetry 
breaking---an open question at present. 

$SU(2)$ breaking brings in an important parameter, 
the ratio
\begin{equation}
\tan\beta=\frac{\langle H_2\rangle}{\langle H_1\rangle},
\end{equation}
which is bounded by the requirement of 
perturbative Yukawa couplings to be approximatively 
between $1$ and $50.$
At the same time, the higgsinos and the gauginos 
(except the gluino)
mix; the resulting mass eigenstates are called 
neutralinos ($\tilde\chi^0_i$) and charginos ($\tilde\chi^+_i$).
Another effect of $SU(2)$ breaking is the mixing
of the squarks which are partner of the left and the right quarks
(called left and right squarks); similarly for the sleptons.
We have: 
\begin{equation}
\begin{array}{rcl}
m^2_{\tilde {u_i}{\LR}}&=&
m_{u_i} (A_{u_i}-\mu^*\ {\rm cot}\beta),\\
m^2_{\tilde {d_i}{\LR}}&=&
m_{d_i} (A_{d_i}-\mu^*\ {\rm tan}\beta),
\end{array}
\label{leftright}
\end{equation}
where $i=1,2,3$ is the generation index, 
$A_{u_i}$ and $A_{d_i}$ are parameters of the Higgs-squark-squark, 
soft-supersymme\-try-breaking interactions and
$\mu$ is a supersymmetry-conserving mass parameter.
These same parameters enter also the interaction vertices
of the Higgs with the squarks, 
as we will see below in a noticeable case.

The massive supersymmetric parameters can  
be complex; in particular phases 
can be present in the gaugino masses, in the $A$- and
$\mu$-parameters 
defined above\fnote{\dag}{In the constrained version 
called {\em low-energy supergravity model} there may be at most 
two new phases.}.
These phases lead to a muon transverse polarization 
even if there is no new flavor violation 
that is related to the supersymmetric parameters; 
our study extends the analysis of \cite{CF} 
to the region in which $\tan\beta$ is large. 
Then we will comment on 
the effect of relaxing the minimal hypothesis
on supersymmetric flavor violation; 
the importance of this point has been stressed in~\cite{WuNg}.

\vskip.3truecm 
{\it The transverse polarization}
\vskip.3truecm 
To evaluate the possibility to have a positive 
experimental signal, 
we need an estimation of the 
muon transverse polarization.
In the following an {\em upper bound} on this effect is given.
It shows that the transverse polarization 
is too small to be detected.

For this sake, let us
consider the gluino exchange diagram in Fig.\ 1(a).
The new supersymmetric phases are present in the
gluino-squark loop, which induces the effective 
coupling between the 
$u$- and $s$-quarks and the charged Higgs field $H^-$:
\begin{equation}
\frac{g_s^2}{(4\pi)^2}\  
\left(V_{us} \frac{m_s}{v}\ A_s  \tan\beta \right)\
\frac{m_{\tilde g}}{m_{\tilde q}^2} 
\cdot (\bar{s}(x)\ P_L\ u(x))\ H^-(x) ,
\label{higgseff}
\end{equation}
where $V$ is the Cabibbo-Kobayashi-Maskawa matrix, 
$g_s$ the strong coupling and $v=174$ GeV.
In  formula \refe{higgseff} we only kept 
the part of the $H^-$-$\tilde q$ coupling that 
grows with $\tan\beta;$
the gluino mass $m_{\tilde g}$ provides the chirality flip, and
the squark mass $m_{\tilde q}$ gives the correct dimension.
The exchange of charged Higgs $H^-$ 
therefore lead to the effective operator:
\begin{equation}
\sin\theta_C\ G_S\cdot 
(\bar{s}(x)\ P_L\ u(x) )\ (\bar\nu_\mu(x)\ P_R\ \mu(x)),
\end{equation}
where, using \refe{scalar}:
\begin{equation}
G_S=\frac{1}{m_{H^-}^2}\
\frac{g_s^2}{(4\pi)^2}\ \frac{m_\mu\ m_s}{v^2}\ 
\frac{A_s\ m_{\tilde g}}{m_{\tilde q}^2}\ \tan^2\beta .
\label{quadratic}
\end{equation}
The coupling of the charged Higgs with the lepton has brought in
a second factor $\tan\beta,$ and this is the origin of the
enhancement of this type of diagram.

Accordingly to previous discussion we find:
\begin{equation}
F_S= \frac{G_S}{2\sqrt{2} G_F}\ \frac{f_S}{f_+} \sim
6\times 10^{-5}\ 
\left(\frac{A_s\ m_{\tilde g}}{m_{\tilde q}^2}\right)\
\left(\frac{100\ {\rm GeV}}{m_{H^-}}\right)^2 \ 
\left(\frac{\tan\beta}{50}\right)^2 .
\label{opt}
\end{equation}
This estimation shows that 
non-zero contributions to the form factor 
$F_S$ are possible in the minimal supersymmetric 
extension of the standard model if the supersymmetric
parameters $A_s$ or $m_{\tilde g}$ are complex.
However, it also shows why it is difficult 
to expect a positive signal in the next generation 
searches of transverse polarization.
In fact,
in the numerical estimation \refe{opt}, 
the transverse polarization can attain relatively
large values only if we assume:
\begin{description}  
\item (a) quite light supersymmetric masses;
\item (b) large values of $\tan\beta$; 
\item (c) large supersymmetric phases 
(in the gluino masses and $A_s$). 
\end{description}
We recall two possible 
unpleasant features of the large $\tan\beta$ scenario:
first, it usually implies fine-tuning in the 
parameters of the scalar 
sector; secondly, 
amplitudes depending on the Yukawa parameters 
(like those for ``dimension five'' proton decay, 
or those for $b\to s\gamma$ transition) 
may become too large in this limiting case.
(Notice, for the following discussion, 
that $m_{\tilde{b}\LR}^2$ defined in \eq{leftright}
is expected to be large, 
of the order of $m_b \times \mu \tan\beta:$ 
cancellations with the $A_b$-term contribution 
would imply color-breaking-minima 
in the scalar potential.)
Furthermore, let us remark that the parameter $A_s$ 
is not free from experimental constraints.
In fact the gluino $s$-quark loop will generate
an electric dipole moment for the $s$-quark; 
this effect is further amplified by a $\tan\beta$ factor
(present in the left-right $s$-squarks mixing).

Let us consider now the diagram in Fig.\ 1(b),
that is obtained from Fig.\ 1(a) by replacing
$u$ with $\nu,$ $s$ with  $\mu,$ and $\tilde g$ with $\tilde Z.$
The loop is at the leptonic end, and the phases 
now appear in the soft-breaking leptonic parameter $A_\mu$ 
and in the zino mass. 
The contribution of this 
diagram has the same quadratic behavior in $\tan\beta$
discussed in \refe{quadratic}.
The amplitude is few times smaller, 
due to the weak gauge coupling replacing the strong one.
However, the limits from muon electric dipole moment 
are much weaker. 
For this reason, the leading effect might be related  
to neutralino exchange diagrams.

In view of the negative result 
we will not proceed in the discussion, and
regard \refe{opt} as an upper bound on the effect.
Such an upper bound is rather robust in the sense that 
the final number is small, no matter what loop diagram 
we consider.
\vskip.3truecm 
{\it On Supersymmetric Flavor 
Violations and Transverse Polarization}
\vskip.3truecm
Recently, a supersymmetric 
scenario which makes room for observable effects
in the next generation of transverse polarization
experiments has been proposed \cite{WuNg}.
The scenario relies on the sources of flavor violation that  
are provided by the supersymmetric parameters: 
this leads to the possibility that the gluino couplings 
with the $u$- and the $s$-quarks involve
the third family squarks, those with the largest
couplings with the charged Higgs 
(compare with eq.\ \refe{higgseff}).
Three major assumptions have to be fulfilled:
\begin{description}
\item (a) $V^{D_L}_{32},$ that quantifies
the mixing of the ``left'' squark  $\tilde s_L$ 
with $b_L$ in the gluino coupling, and
$V^{U_R}_{31},$ analogously defined, are order unity 
and carry large phases;
\item (b) the masses of the gluino and of the squarks
are close to their present experimental value;
\item (c) $\tan\beta$ is large.
\end{description}
Unfortunately, this set of assumptions becomes problematic
as soon as we consider the rate of the 
$b\to s\gamma$ transition, that is known to be fairly well reproduced 
by the standard model amplitude alone.

To make this point explicit, let us consider the 
gluino-bottom-squark diagram for this transition. 
The gluino mass provides the chirality flip, 
and one $m^2_{\tilde{b}\LR}$-insertion in the bottom-squark line
allows one to construct the dipole operator. 
Let us compare this contribution with that of the  
standard model:
\be
\frac{{\cal M}_{\tilde g}}{{\cal M}_W}
=
\frac{  \frac{e \alpha_s}{4\pi}\ 
        \frac{m_{\tilde g}}{m_{\tilde b}^2}\
        (V_{32}^{D_L}\ \frac{m_{\tilde{b}\LR}^2}{m_{\tilde b}^2}\ 
        V_{33}^{D_R})\
        {\cal F}_{\tilde g}
               \left(\frac{m_{\tilde b}^2}{m_{\tilde g}^2}\right)}
        {\frac{e \alpha_W}{4\pi}\ 
        \frac{m_b}{m_W^2}\
        (V_{ts} V_{tb})\
        {\cal F}_W\left(\frac{m_t^2}{m_W^2}\right)} .
\l{estim}
\ee 
The loop functions ${\cal F}$ were computed in \cite{Stefano};
assuming, consistently with \cite{WuNg}, that $m_{\tilde b}\sim   
m_{\tilde g},$ the functions amount to a factor close to unity.
{}From \eq{estim} we come to the estimate:
\be
\frac{{\cal M}_{\tilde g}}{{\cal M}_W}
\sim 
10^3\times  
\left( \frac{V_{32}^{D_L}}{1/\sqrt{2}}\right)\  
\left( \frac{V_{33}^{D_R}}{1/\sqrt{2}}\right)\  
\left( \frac{\tan\beta}{50}\right)\  
\left( \frac{\mu\ m_{\tilde g}\ m_W^2 }{m_{\tilde b}^4}\right)
\ee
where we assumed $V_{ts}\sim V_{cb}.$
Therefore this contribution may 
trigger a $b\to s\gamma$ transition one million time faster
then that of the standard model.

The way out suggested in~\cite{WuNg}
is that the chargino contribution, whose importance 
has been emphasized in the literature \cite{cancellation}, 
cancels the gluino amplitude and accordingly
makes the bounds discussed above less stringent.
It is however hard to justify such a precise 
cancellation, since different 
parameters enter the two amplitudes.
Moreover, it is not clear if it is possible to implement
such a cancellation without
suppressing the transverse polarization effect as well.
Finally, in a purely phenomenological 
scenario, like the one considered, one would expect the dominance 
of the gluino exchange amplitude over all the other ones,
chargino exchange included, since no smallness factors 
are attached to the gluino 
couplings\fnote{\dag}{A different situation happens
in the models where flavor-changing gluino couplings are 
induced radiatively
by the usual Yukawa couplings, and therefore are
strongly suppressed---see for instance \cite{Stefano}.}.

In conclusion, we feel that it is difficult 
to account for a transverse
polarization within the 
context of the minimal supersymmetric extension of the
standard model even after making allowence for family mixing and
flavor violations. 

\subsection{R-parity Breaking Models}
\vskip.3truecm 
{\it Introduction}
\vskip.3truecm 
The requirement of gauge invariance permits the following
renormalizable interactions in the superpotential:
\be
\begin{array}{l}
(Y^E_{jk}\ H_1 +\lambda_{ijk}\ L_i)\ L_j\ E^c_k 
+ (Y^D_{jk}\ H_1 +\lambda'_{ijk}\ L_i)\ Q_j\ D^c_k\\ 
+ (\mu\ H_1+\mu_i\ L_i)\ H_2-
Y^U_{jk}\ H_2\ Q_j\ U^c_k + \lambda''_{ijk}\ D^c_i\ D^c_j\ U^c_k.\\
\end{array}
\label{superpot}
\ee
Besides the interactions of quark ($Q,\; U^c,D^c$) and 
lepton  ($L,E^c$) with Higgs ($H_1, H_2$) superfields, and the 
$\mu$-term, we have the {\em R-parity breaking} interactions,
parameterized by $\lambda,\lambda',\lambda''$ and $\mu_i,$ that
have no correspondence in the standard model 
lagrangian, and break either
the lepton ($\lambda,\lambda',\mu_i$) or the baryon ($\lambda''$) 
number.
We consider strict baryon number conservation ($\lambda''=0$) 
in the following, in order 
to avoid strong matter stability bounds \cite{CRS,Alexei}.

Since the R-parity breaking couplings are {\em a-priori}
complex quantities, as remarked in~\cite{Liu}, 
it is important to ask whether they can 
manifest themselves in a large $\pt.$ 
Before answering this question
we must recall some
relevant information on the model under consideration.

Let us assume a generic pattern of the R-parity
breaking couplings $\lambda$, $\lambda'$, and $\mu_i.$ 
We fix the basis in the $H_1$, $L_i$ 
four-dimensional space in two steps: 
\begin{description}
\item (1) we redefine the Higgs superfield
in such a way that $\mu_i$ terms are absent;
\item (2) we further 
rotate the three lepton superfields in order 
to make the lepton mass matrix diagonal (analogously for
the quark superfields).
\end{description}

In general, R-parity breaking interactions of scalars induce 
vacuum expectation values of sneutrino fields 
$\tilde{\nu}_i$ \cite{Hall-Suzuki,Lee}.
In fact the similarity between 
the scalar leptons and the usual 
Higgs doublets is almost complete 
in the model under consideration.

Supersymmetry manifests itself by
providing relations among the various 
interactions; for instance, 
$\lambda'_{ijk}\ L_i\ Q_j\ D^c_k$
describes at the same time the 
interactions of quarks with   
the slepton $\tilde{l}_i(x)$
(that in this context can be thought of
as an Higgs doublet)
and the interactions
in which the squarks behave as leptoquarks.  
This implies that this model is more 
predictive, and more 
constrained, than a multiHiggs (or a leptoquark) model.

The Yukawa $Y^D, Y^E$ interactions are fixed 
by the tree level condition: 
\be
\begin{array}{l}
Y^D_{jk}=\frac{M^D_{jk}}{\langle H_1\rangle} 
- \lambda_{ijk}'\ 
\frac{\langle\tilde{\nu}_i\rangle}{\langle H_1 \rangle} \\[2pt] 
Y^E_{jk}=\frac{M^E_{jk}}{\langle H_1\rangle} 
- 2\lambda_{ijk}\ 
\frac{\langle\tilde{\nu}_i\rangle}{\langle H_1 \rangle} 
\end{array}
\ee
to reproduce the observed fermion masses.
The $\lambda$ and $\lambda'$ couplings 
are also subject to experimental bounds from 
of various processes \cite{CRS,bounds}.
In particular the couplings that trigger 
flavor-changing neutral current
transitions are quite strongly constrained.

Neutrino masses can be induced 
in this model by two mechanisms: due to the mixing of neutrinos
and zinos, caused by the sneutrino vacuum expectation values,
but also by fermion-scalar loops with two R-parity breaking vertices
\cite{Hall-Suzuki}. 
We will be interested in this second type of contribution in the
following. 
Let us therefore recall that the coupling $\lambda'_{ijj}$ gives: 
\be
(\delta m_{\nu_i})^{\rm\scriptscriptstyle loop}
\sim\frac{3}{8\pi^2} \
\frac{(\lambda'_{ijj})^2}{m^2_{\tilde{d}_j}}\ 
m_{d_j}\, m^2_{\tilde d_j\LR} 
\l{neuloop}
\ee
where the factor 3 is for color; this factor 
is absent for loops induced by $\lambda$ couplings. 
A glance at eq.\ \refe{leftright} shows that  
there are two fermion mass insertions in \refe{neuloop}:
this must be so, since the gauge invariant operator 
for neutrino mass is of the form 
neutrino-neutrino-Higgs-Higgs, 
and the Higgs field is coupled to fermion masses.
\vskip.3truecm 
{\it The transverse polarization}
\vskip.3truecm 
Let us assume that the couplings in the interactions:
\be   
   \lambda_{322}'\ L_3\ Q_2\ D^c_2\ \ \ \ {\rm and}\ \ \ \  
   \lambda_{322}\ L_3\ L_2\ E^c_2
\l{neededcoupl}
\ee
in \refe{superpot} 
are not small, whereas the other R-parity breaking couplings 
are suppressed to obey the experimental bounds. 
Let us further assume that the quark doublet  
is $Q=(V^\dagger\ U,\ D),$ in order to avoid sneutrino mediated
flavor-changing interactions.
The couplings in eq.\ \refe{neededcoupl} 
induce, after integrating away the slepton $\tilde{e}_3,$
the effective operator:
\be
\sin\theta_C\ G_S\cdot 
(\bar s(x)\ P_L\ u(x) )\ (\bar\nu_\mu(x)\ P_R\ \mu(x)),
\ee
where 
\be
G_S= \frac{\lambda'_{322}\ \lambda_{322}}{m^2_{\tilde{e}_3}}.
\ee
This yields the scalar form factors:
\be
F_S= \frac{G_S}{2\sqrt{2} G_F}\ \frac{f_S}{f_+} \sim
4\times 10^{-2}\ \left(\frac{\lambda_{322'}}{0.1}\right) 
\left(\frac{\lambda_{322}}{0.1}\right)
\left(\frac{100\ {\rm GeV}}{m_{\tilde{e}_3}}\right)^2. 
\ee
Comparing with eqs.\ \refe{xi} and \refe{obs}, we conclude that
a phase larger than $1/20$ 
would lead to a positive signal in the next generation 
searches of transverse polarization

The couplings above are however subject to bounds due 
to the neutrino masses.
Using the estimate in eq.\ \refe{neuloop}, and asking 
the {\em tau neutrino} mass to be 10 eV, we find an upper limit
on the couplings considered:
$\lambda_{322}\ltap 0.02$ and 
$\lambda_{322}'\ltap 0.03$,  
where the supersymmetric massive parameters
have been assumed around 100 GeV. This implies:
\be
\Im\xi_S\ltap 2\times 10^{-3} ,
\l{neubound}
\ee
that, with a phase of order unity 
could still give an observable transverse polarization effect.
On the contrary, if the scale of supersymmetry breaking 
would be one order of magnitude larger,
around the TeV, it would be 
improbable to have a detectable effect.
Let us notice that a neutrino in the 10 eV range
is a hot dark matter candidate, and 
therefore well motivated.
There are several possibilities that allow one to relax the
bound \refe{neubound}
(but, to advocate one or more of them, 
would take us beyond the present 
phenomenological approach). One can:
\begin{description}
\item (1) assume a cancellation in the left-right 
mixing in eq.\ \refe{neuloop}
to diminish the induced neutrino mass;
\item (2) allow for compensation between the
$\lambda$ and $\lambda'$ contributions, or else
partial cancellations of the loop-induced and
the sneutrino-vacuum expectation value 
contributions to $m_{\nu_\tau}$;
\item (3) or, finally, impose the weaker bound 
coming 
from experimental studies of tau decays,
$m_{\nu_\tau}\ltap 20$ MeV.
\end{description}
Last possibility can be regarded with 
favor, since it does not rely on cancellations.
Accepting this option,
one should however bear in mind that a
neutrino heavier than approximatively 100 eV has to
be unstable to avoid cosmological bounds.
In view of the above mentioned difficulties, and
also of the very specific choice of couplings
in eq.\ \refe{neededcoupl}, we conclude that 
there is only a marginal possibility 
that an observable transverse polarization is 
related to this kind of models. 

\subsection{Other Models}

Models that can generate a sizeable 
transverse polarization has been discussed in the literature.
In particular, models including
leptoquarks or new Higgs particles
(more in general: new scalars coupled
to fermions) \cite{Zhitniskii,Cheng,Leurer,CFK,gk,bg}, 
or also fundamental tensor particles \cite{Chizhov}
has been considered. 

Models with new scalars are probably 
the most promising candidates to account for a positive signal. 
In fact, complex part of the form factor of the order of:
\beq
\Im\xi_S \sim  8 \times 10^{-3} \, ,
\eeq
can give rise to a measurable muon transverse polarization 
without conflicting with other observables.

\section{Conclusions}

The knowledge of the form factors in the $K^0_{\mu 3}$ process 
could give important information on CP-violation.  
The experimental signature is provided
by the transverse polarization of the muon. 

At the level of a purely phenomenological analysis, 
we stressed the complementarity of this experimental information 
with that from the analysis of the Dalitz-plot distribution; 
we also pointed out the importance of distinguishing between 
scalar and tensor form factors.

The discussion of realistic models was 
focused on the supersymmetric extensions of the standard model. 
In the minimal model, assuming that the Cabibbo-Kobayashi-Maskawa 
matrix is the only source of flavor violation,
the contribution to the muon polarization 
is too small, even under optimistic 
assumptions, as in \refe{opt}. 
We commented on the scenario in which large
supersymmetric CP and flavor violations are allowed,
and showed the difficulties that arise of reconciling 
the assumptions required to have a large $\pt$ and
the observed rate of the $b\to s\gamma$ transition. 
Observable effects are in principle 
possible departing from the hypothesis of R-parity conservation,
but constraints from the neutrino masses 
severely restricts 
the region of parameter space
in which this can happen.
 
In conclusion, assuming that the 
minimal supersymmetric standard model
(possibly with R-parity breaking interactions)
is correct, we expect that future search of
transverse polarization in $K^0_{\mu 3}$ decay should give a 
null result. 
On the other hand, a signal of transverse 
polarization in forthcoming experiments would point to
physics different from the standard model and from its
straightforward supersymmetric extensions.

\vskip1truecm
{\sc Acknowledgements}
\vskip0.5truecm 
\noindent The authors would like to thank 
S.\ Bertolini and E.\ Christova for discussions. 
F.\ V.\ acknowledges conversations with 
M.\ Chizhov,  M.A.\ Diaz, 
N.\ Paver, A.\ Ra\v{s}in,
G.\ Senjanovi\'c and A.Yu.\ Smirnov.

\newpage


\begin{figure}[t]

\begin{center}

\begin{picture}(200,100)

\put(93,-30){(a)}

\put(0,100){\line(1,0){200}}         
\put(50,100){\vector(1,0){0}}
\put(50,110){$\mu_R$}
\put(150,100){\vector(1,0){0}}
\put(150,110){$\nu_L$}

\multiput(100,33.3333)(0,10){7}{\line(0,1){5}}
\put(100,66.666){\vector(0,-1){0}}
\put(105,60){$H^-$}

\multiput(50,0)(18,12){3}{\line(3,2){10}}
\put(75,16.6666){\vector(3,2){0}}
\put(60,20){$\tilde u_L$}

\multiput(100,33.333)(18,-12){3}{\line(3,-2){10}}
\put(125,16.6666){\vector(3,-2){0}}
\put(130,20){$\tilde s_R$}

\put(0,0){\line(1,0){50}}         
\put(25,0){\vector(1,0){0}}
\put(25,-10){$u_L$}

\put(150,0){\line(1,0){50}}       
\put(175,0){\vector(1,0){0}}
\put(175,-10){$s_R$}

\put(50,0){\line(1,0){100}}
\put(100,-10){$\tilde g$}

\end{picture}

\vskip3.5truecm

\begin{picture}(200,100)

\put(93,-25){(b)}

\put(0,0){\line(1,0){200}}         
\put(50,0){\vector(1,0){0}}
\put(50,-10){$u_L$}
\put(150,0){\vector(1,0){0}}
\put(150,-10){$s_R$}

\put(0,100){\line(1,0){50}}         
\put(25,100){\vector(1,0){0}}
\put(25,110){$\mu_R$}

\put(150,100){\line(1,0){50}}       
\put(175,100){\vector(1,0){0}}
\put(175,110){$\nu_L$}

\put(50,100){\line(1,0){100}}
\put(100,110){$\tilde{Z}$}

\multiput(50,100)(18,-12){3}{\line(3,-2){10}}
\put(75,83.33333){\vector(3,-2){0}}
\put(60,75){$\tilde \mu_R$}

\multiput(100,66.6666)(18,12){3}{\line(3,2){10}}
\put(125,83.33333){\vector(3,2){0}}
\put(130,75){$\tilde \nu_L$}

\multiput(100,0)(0,10){7}{\line(0,1){5}}
\put(100,35){\vector(0,-1){0}}
\put(105,35){$H^-$}

\end{picture}

\end{center}

\vskip0.7truecm

\caption{Two Feynman diagrams 
inducing scalar-type four-fermion interactions 
in the minimal supersymmetric 
extension of the standard model.}

\end{figure}


\newpage

\begin{thebibliography}{99}

\bibitem{Sakurai}
J.J. Sakurai, {\em Phys. Rev.} {\bf 109} (1958) 980.

\bibitem{Zhitniskii}
A.R. Zhitnitskii, {\it Yad. Fiz.} {\bf 31} (1980) 1024
[\sjnp{31}{80}{529}].

\bibitem{Golowich}
E. Golowich and G. Valencia, \prd{40}{80}{112}.

\bibitem{DAFNE}
``The DA$\Phi$NE Handbook'', 
eds. L. Maiani, G. Pancheri, N. Paver,
Frascati 1992.

\bibitem{Kuno} Y. Kuno, {\em Nucl. Phys. B} (Proc. Suppl.) 
{\bf 37A} (1994) 87.

\bibitem{BNL}
R. Adair \etal, 
``Muon Polarization Working Group Report'',
{\tt hep-ex/9608015.}

\bibitem{old}
S.R. Blatt \etal, \prd{27}{83}{1056}.

\bibitem{Privitera}
P. Privitera, 
``Measurement of Muon Polarization 
in \kmu3\ at a $\phi$ Factory'',
{\tt hep-ph/9605416.}

\bibitem{Cheng}
H.Y. Cheng, \prd{26}{82}{143}.

\bibitem{Leurer}
M. Leurer, \prl{62}{89}{1967}.

\bibitem{CFK}
P. Castoldi, J.-M. Fr\`{e}re and G.L. Kane, \prd{39}{89}{2633}.

\bibitem{gk} R. Garisto and G. Kane, \prd{44}{91}{2038}.

\bibitem{bg} G. B\`elanger and C.Q. Geng, \prd{44}{91}{2789}.

\bibitem{CF} E. Christova and M. Fabbrichesi, \plb{315}{93}{113}. 

\bibitem{WuNg}
G.-H. Wu and J.N. Ng, 
``Supersymmetric Time Reversal Violation 
in Semileptonic Decays of Charged Mesons'',
{\tt hep-ph/9609314.}

\bibitem{PDG}
Particle Data Group, \prd{54}{96}{1};\\
{\em idem}, \plb{111}{82}{73}. 

\bibitem{OK}
L.B. Okun and I.B. Kriplovich, {\it Yad. Fiz.} {\bf 6} (1967) 821
[\sjnp{6}{68}{598}].

\bibitem{pt1}
S.W. MacDowell, \nc{9}{58}{258}.

\bibitem{pt2}
N. Cabibbo and A. Maksymowich, {\em Phys. Lett.} {\bf 9} (1964) 352;
Erratum {\em ibidem} 11 (1964) 360; 
{\em ibidem} 14 (1966) 72. 

\bibitem{Akimenko}
S.A. Akimenko \etal, {\em Phys.Lett.} {\bf B259} (1991) 225.

\bibitem{OkunBook} 
``Leptons and Quarks'', L.B. Okun, 
Amsterdam 1982, North-holland.

\bibitem{BP} J. Bijnens, J. Prades and E. de Rafael, 
\plb{348}{95}{226}.

\bibitem{Chizhov}
M. Chizhov, \mpla{8}{93}{2753}.\\ 
A subsequent work, \plb{381}{96}{359} is specifically dedicated 
to the  study of some aspects of tensor interactions at DA$\Phi$NE.

\bibitem{QM} K. Nishijima, \nc{11}{59}{698};\\
F. Gursey, \nc{16}{60}{230} and \ap{12}{61}{91};\\
J.A. Cronin, {\it Phys. Rev.} {\bf 161} (1967) 1483;\\
S. Weinberg, {\it Physica} {\bf  96A} (1979) 327;\\
A. Manohar and H. Georgi, \npb{234}{84}{189};\\
A. Manohar and G. Moore, \npb{243}{84}{55};\\
D. Espriu, E. de Rafael and J. Taron, \npb{345}{90}{22}. 

\bibitem{Stefano}
S. Bertolini, F. Borzumati, A. Masiero and G. Ridolfi,
\npb{353}{91}{591}.

\bibitem{cancellation} R. Barbieri and G. F. Giudice, 
\plb{309}{93}{86},
discussion of the cancellation in the supersymmetric limit;\\
J. Lopez, D. Nanopoulos and G. Park, 
\prd{48}{93}{974}, dependence on the sign of $\mu$;\\ 
N. Oshimo, \npb{404}{93}{20} and 
M. Diaz, \plb{322}{94}{207},
dependence on $\tan\beta$;\\
Y. Okada, \plb{315}{93}{119}, dependence on the
stop splitting;\\
F. Borzumati, \zpc{63}{94}{291} and
P. Nath and R. Arnowitt, \plb{336}{94}{395},
discussion of the cancellation based 
on a numerical analysis of the rate;\\
R. Garisto and J.N. Ng, \plb{315}{93}{372} and
S. Bertolini and F. Vissani, \zpc{67}{95}{513},
discussion based on the analytical form of the chargino amplitude.

\bibitem{CRS}
C.E. Carlson, P. Roy and M. Sher,
\plb{357}{95}{99}.

\bibitem{Alexei}
A.Yu. Smirnov and F. Vissani,
\plb{380}{96}{317}.

\bibitem{Liu}
C. Liu, \ijmpa{11}{96}{4307}.

\bibitem{Hall-Suzuki}
L.J. Hall and M. Suzuki, \npb{231}{84}{419}.

\bibitem{Lee} I-H. Lee, \plb{138}{84}{121}; \npb{246}{84}{120}.

\bibitem{bounds}
V. Barger, G.F. Giudice and T. Han,
\prd{40}{89}{2987};\\
K. Agashe and M. Graesser, \prd{54}{96}{4445};\\
F. Vissani, 
``R-Parity Breaking Phenomenology'', {\tt hep-ph/9602395;}\\
D. Choudhuri and P. Roy, \plb{378}{96}{153};\\
G. Bhattacharyya, 
``R-Parity Violating Supersymmetric Yu\-ka\-wa
Couplings: A Minireview'', {\tt hep-ph/9608415.}

\end{thebibliography}
\end{document}